\documentclass{aa}
\usepackage{graphicx}
\usepackage{txfonts}
\usepackage{natbib}

\begin{document}

\title{Dynamics of small-scale convective motions}

\author{B. Lemmerer
\inst 1
\and A. Hanslmeier
\inst 1
%\inst 2
\and H. Muthsam
\inst 2
\and I. Piantschitsch
\inst 1
}

\institute{Institute of Physics, IGAM, University of Graz, Universit{\"a}tsplatz 5, 8010 Graz, Austria
\and Institute of Mathematics, University of Vienna, Oskar-Morgenstern-Platz 1, 1090 Wien, Austria
}

\date{received 21 December 2015 / accepted 4 October 2016}

\abstract{Previous studies have discovered a population of small granules with diameters less than 800 km located in the intergranular lanes showing differing physical properties. High resolution simulations and observations of the solar granulation, in combination with automated segmentation and temporal tracking algorithms, allow us to study the evolution of the structural and physical properties of these granules and surrounding vortex motions with high temporal and spatial accuracy.}{We focus on the dynamics of granules, that is, the lifetime of granular cells, the fragmentation behavior, the variation of size, position, emergent intensity and vertical velocity over time and the influence of strong vortex motions. Of special interest are the dynamics of small granules compared to regular-sized granules.}{We developed a temporal tracking algorithm based on our previously developed segmentation algorithm for solar granulation. This was applied to radiation hydrodynamics simulations and high resolution observations of the quiet Sun by SUNRISE/IMaX.}{The dynamics of small granules differ in regard to their diameter, intensity and depth evolution compared to the population of regular granules. The tracked granules in the simulation and observations reveal similar dynamics regarding their lifetime, evolution of size, vertical velocity and intensity. The fragmentation analysis shows that the majority of granules in the simulations do not fragment, while the opposite was found in the observations. Strong horizontal and vertical vortex motions were detected at the location of small granules. Compared to granules, regions of strong vertical vorticity show higher intensities and higher downflow velocities, and live up to several minutes.}{The analysis of granules separated according to their diameter in different groups reveals strongly differing behaviors. The largest discrepancies can be found within the groups of small, medium-sized and large granules. Therefore, these groups have to be analyzed independently. The predominant location of vortex motions on and close to small granules indicates a strong influence on the dynamics of granules.}


\maketitle

\section{Introduction}
\label{sec:Introduction}
The study of the dynamics of solar granulation has come back into focus with the production of new high-resolution data from recently installed telescopes (NST, SST, GEGOR) or satellites (Hinode). High resolution observations are crucial for studying structural and physical properties of small-scale convective features such as granulation. Granulation has been investigated by observations \citep[see, e.g.,][]{1986SoPh..107...11R,1989ApJ...336..475T,1990ARA&A..28..263S,1999ApJ...515..441H,2001A&A...377L..14R,2003AN....324..405B} and simulations \citep[see, e.g.,][]{1985SoPh..100..209N,1989ApJ...342L..95S,1989A&A...213..371S,2000A&AS..146..267G}. Statistics can only be calculated efficiently using methods for the automated segmentation of the solar granulation. Development of these methods started in the early 90s  \citep[see, e.g.,][]{1989ApJ...336..475T,1986SoPh..107...11R,1997ApJ...480..406H}. The relatively low spatial resolution was a limiting factor for the achievable accuracy of the algorithms. The segmentation was mainly reduced to applying single threshold segmentation routines or computationally cheap edge-detection methods to observations. \citet{2001SoPh..201...13B} introduced an algorithm that provides a powerful method for automated granule detection on intensity maps, based on the combined segmentation on multiple intensity thresholds. In \citet{2014A&A...563A.107L} (hereafter `paper I'), we introduced a segmentation algorithm that is based on both intensity and velocity information. The incorporation of velocity images in the method leads to an improvement in the quality of the segmentation of granules and also helps to detect the less bright upwards or downwards moving cells.\\

The increased accuracy of the segmentation combined with higher resolution data allows a finer distinction in the analysis of the granulation. This study focuses on a distinct sub-population of small granular cells with diameters less than 600 km first detected by \citet{2012ApJ...756L..27A} studying observational data of the New Solar Telescope (NST) at Big Bear Solar Observatory (BBSO). In Paper I these findings were found to be in accordance with the analysis of simulated data using snapshots of the ANTARES simulation, a radiation hydrodynamics simulation of the solar photosphere and convection zone with high spatial and temporal resolution. We found that small granules with diameters below 800 km have a similar structural appearance as regular sized granules but show different dynamics. These small cells are formed in clusters between large granules and appear less bright, show smaller vertical velocities and are located deeper into the upper convection zone. Hence, we concluded that the small granules form a separate population of granules. In this study the granules are grouped according to their maximum diameter in their evolution:
\begin{itemize}
\item Small granules: diameter less than 800 km.
\item Medium-sized granules: diameter between 800 km and 1700 km.
\item Large granules: diameter larger than 1700 km.
\end{itemize}

Here, we continue our research on granules in high resolution simulations and observations. High-cadence time-series of intensity maps show that small granules seem to appear on the solar surface between medium-sized granules and disappear again by sinking back without fragmenting but instead keeping their size. To study this process in detail we focus on the dynamics of granules, that is, their lifetime, fragmentation process, and vicinity,  providing us with useful information on processes that effect their appearance on the visible surface. Hence, we developed a temporal tracking algorithm based on our segmentation algorithm (see Paper I). For each segmented granule, different properties can be extracted and tracked, such as the size, the position in the simulation box, the emergent intensity and the vertical velocity. The tracking algorithm is applied to high resolution radiation-hydrodynamics (RHD) simulations and observations of the quiet Sun. The application of the algorithm helps to quantify the visual interpretation of temporal series from simulations and observations.\\ 

Organization of this paper: In section 2 we introduce the numeric simulation code ANTARES \citep[see] []{2010NewA...15..460M} that provides the simulated profiles of the emergent intensity and vertical velocity on which our segmentation and tracking routine is based. We compare the study on the simulated profiles to high resolution images of the quiet Sun provided by the SUNRISE telescope. The pre-processing of the data and the newly developed tracking algorithm are described in section 3. In section 4 we present the results of the tracking of granules in simulations and observations. Additionally, we focus on the dynamics of strong vortex motions and their influence on granules in simulations. The final section presents our discussion and conclusions as well as our outlook on future studies.

\section{Data and simulation setup}
\label{sec:Data}
\subsection{Numerical simulations}
The ANTARES (A Numerical Tool for Astrophysical RESearch) code is a numeric simulation code of the solar near-surface convection \citep[“box-in-a-star” approach; see][]{2010NewA...15..460M}. The code solves the set of radiation hydrodynamic (RHD) equations using weighted, essentially non-oscillatory (WENO), high-resolution numerical schemes \citep[see] []{2013A&A...554A.119Z,2013MNRAS.435.3191M}. In the vertical direction, open boundary conditions allow free in- and outflow, while in the horizontal directions periodic boundary conditions are used \citep[for a detailed description of the boundary conditions, see] []{2014CoPhC.185..764G}.\\

As an initial condition for the model, a simulation box 6 Mm wide and 4.5 Mm deep is used. This is copied three times in each horizontal direction resulting in a horizontal extent of 18 Mm. The simulation box comprises 18 Mm in the horizontal direction and 4.5 Mm in the vertical direction, with a pixel resolutions of 35 km  and 11 km, respectively. In the upper 1 Mm of the domain, the radiative transfer equation is solved using non-gray opacities with 4 bins. The angular integration is implemented according to a Gauss-Radau formula with 18 rays \citep{2012ApJ...759..120T} and hyperbolic terms are integrated using a 5th order WENO scheme \citep{2012JCoPh.231.3561K}. For the diffusive and viscous terms a 4th order accurate finite-difference approximation is used \citep{2013JCoPh.236...96H}. The model uses the boundary conditions BC 3b from \citet{2014CoPhC.185..764G} and a Smagorinsky subgrid model \citep{1963MWRv...91...99S}.\\ 

\subsection{SUNRISE observations}
Additionally, we analyzed IMaX \citep[Imaging MagnetographeXperiment][]{2011SoPh..268...57M} data obtained by the magnetograph onboard the SUNRISE balloon-borne telescope \citep[see][]{2010ApJ...723L.127S, 2011SoPh..268....1B,2011SoPh..268..103B} operated in 2009 and reaching a height of 37 km above sea level.\\ 

The segmentation and tracking algorithms were applied to data obtained on the 9th of June 2009 between 00:36 UT and 2:02 UT close to the disc center. The data consists of a total of 100 exposures and is divided in two data sets (from 00:36 until 00:59 UT and from 1:31 UT until 2:02 UT). The temporal cadence is approximately 32 secs and the FOV has a size of 50 by 50 $arcsecs^2$ and 936 by 936 $pixel^2$, which was reduced to a size of 780 by 780 $pixel^2$ due to image reconstruction \citep[see][]{2011SoPh..268...57M}. The data was recorded in the V5-6 mode, where the neutral iron line at 525.02 nm was sampled five times in all four Stokes parameters. One polarimetric image is formed by accumulating six exposures resulting in a spatial resolution of approximately 0.15–0.18 arcsec. \\

For our segmentation of the granulation and the analysis we used the line-of-sight (LOS) velocities from the application of the SIR inversion code \citep[see][]{1992ApJ...398..375R} to spectropolarimetric data. For information on the inversion of this data set we refer to \citet{2014ApJ...796...79U}.

\section{Automated tracking of the solar granulation in 2D}
\label{sec:Tracking}
Tracking of the granulation is essential to gain information about the lifetime of granular cells, the fragmentation behavior, the variation of size, position in the simulation box, emergent intensity and vertical velocity over time and the influence of strong vortex motions.\\

In the past, various authors developed different routines for the tracking of the solar granulation. The local correlation tracking  \citep[see, e.g.,][]{1988ApJ...333..427N} or the tracking of individual granules  \citep[see, e.g.,][]{1995ESASP.376b.213S} aim to compute the horizontal flow fields. Based on these two methods new tracking routines have been developed, such as the coherent structure technique \citep{1999A&A...349..301R} that used granules as tracers for tracking the displacement of the center of detected granules. The calculation of the auto-correlation function of a time series \citep[see, e.g.,][]{1989ApJ...336..475T} is a method for determining the time scale of the granulation pattern without previous segmentation of the granulation. \citet{1997ApJ...480..406H} developed a structure tracking method to study the lifetime and evolution of individual granules. The TST \citep[Two-level Structure Tracking;][]{2004A&A...428.1007D} also combines the recognition of granules in single images and their tracking over time. These two methods compare the shape of the detected granules with the original structure within a small region. The methods enable the study of the lifetime of individual granules from the point the granule evolves, up until it dies. A different approach to deriving the lifetime of granules is the calculation of time-slice diagrams \citep[see, e.g.,] []{1999ASPC..183..443R,1998sce..conf..261P,2001SoPh..203..211M}. The position of the prior detected intergranular lanes are followed in a time-series of thin-slit filtergrams. The skeletal representation of the intergranular lanes is then used to classify the granules according to their signatures.\\ 

These methods differ fundamentally in the underlying identification of the granules, which is influenced by the quality of the data (spatial and temporal resolution of the time series) and the purpose and focus of the analysis (see \citet{2014A&A...563A.107L} on the different segmentation techniques). To overcome the difficulties in the identification of granules, we developed a segmentation method that uses multiple intensity thresholds based on different spatial resolutions. This method can be applied to data independent from their spatial resolution and is therefore suitable for observations and simulations. In contrast to the afore mentioned tracking techniques that take all detected granules into account, we divide the detected granules into different groups according to their size. We also focus on the newly discovered population of small granules and compare the lifetime and evolution of their structural and physical properties to medium-sized and large granules.

\subsection{Pre-processing}
The temporal tracking algorithm developed is based on the tracking of separate granules over time using segmentation masks of each time step as input images, which can be derived by applying the 2D segmentation algorithm (described in Paper I) on time series of data from observations and simulations. For each time step, the segmentation algorithm produces a binary mask, which is further used and applied to different physical profiles to analyze, for example, the vertical velocities, the emergent intensities and the vertical position (depth) of the cells  in the simulation box. For each segmentation mask, information on the size, the eccentricity, the coordinates in the image, the mean and maximum physical quantity and the centroid of the segment is stored. The output files form the basis of the tracking routine which accesses the stored information and appends information about detected temporal connections.\\

The tracking algorithm is also applied to SUNRISE IMaX data, which requires the 2D segmentation algorithm to be applied to the Level 2 data, the continuum, and the inverted line-of-sight velocities. Analogous to the analysis of the simulation data, we applied a histogram equalization to the SUNRISE data. We omitted the granules that are located at the image borders. The pre-processing, plus the labeling of all objects, results in a segmentation mask. These masks are then applied to the images to extract statistical information about the identified granules.

\subsection{Tracking algorithm}
The development of a tracking routine, where each granule is traced from the point of appearance to the point of dissolution, requires a data type, allowing us to map the tracking process and to illustrate the fragmentation of cells. These conditions are best met by a `tree' data structure which was therefore implemented. A tree is a hierarchical data structure consisting of:
\begin{itemize}
\item Parent: stores the index of the parent node of the respective tree object.
\item Node: stores the content of the respective tree object. 
\end{itemize}
The root node of a new tree is created if a cell with no parent cell appears in an image. A new node is added to the root node if the detected object in the subsequent time step overlaps with the coordinates of the object in the first image (root node). \\

Our tracking algorithm is a recursive routine, which creates trees by parsing through a temporal series of images. The tree structure displays the evolution of each cell. Due to low intensity values, the segmentation routine sometimes produces errors which result in an insufficient separation of cells. Different plausibility checks are performed to prevent loss of information. The recursive routine is aborted if
\begin{itemize}
\item the overlapping cell in the subsequent image is already part of a tree,
\item the coordinates of a cell overlap with an empty region in the three consecutive images,
\item the cell grows by more than 50\% in the consecutive image and keeps this increased size in the three following consecutive images (this signals the formation of a new cell),
\item the time series ends with the subsequent image.
\end{itemize}
A cell is labeled as `child' of another cell in the subsequent time step if;
\begin{itemize}
\item more than 50\% of the segmented pixels of the cells overlap,
\item the examined cell is not part of a different tree, which means that it is not a child of a different cell,
\item the ratio of the diameter of the consecutive cell and the previous cell which depend on the temporal resolution (simulations: 9 sec, observations: 30 secs) is in the range of $\pm{10}\%$ (for simulations) or $\pm{30}\%$ (for observations).
\end{itemize}
These conditions and restrictions lead to the formation of trees that illustrate the evolution of convective cells. \\ 

The high spatial and temporal resolution of the simulations ensures a consistency in the segmentation and tracking result at each time step. Due to the restricted horizontal extent of the grid, all granules have to be considered by specifically handling the periodic boundaries. In the case of observational data, the field of view is larger and the number of detected cells is higher, therefore granules located at the borders can be neglected.\\ 

\begin{figure*}
        \centering
                \includegraphics[width=1\textwidth]{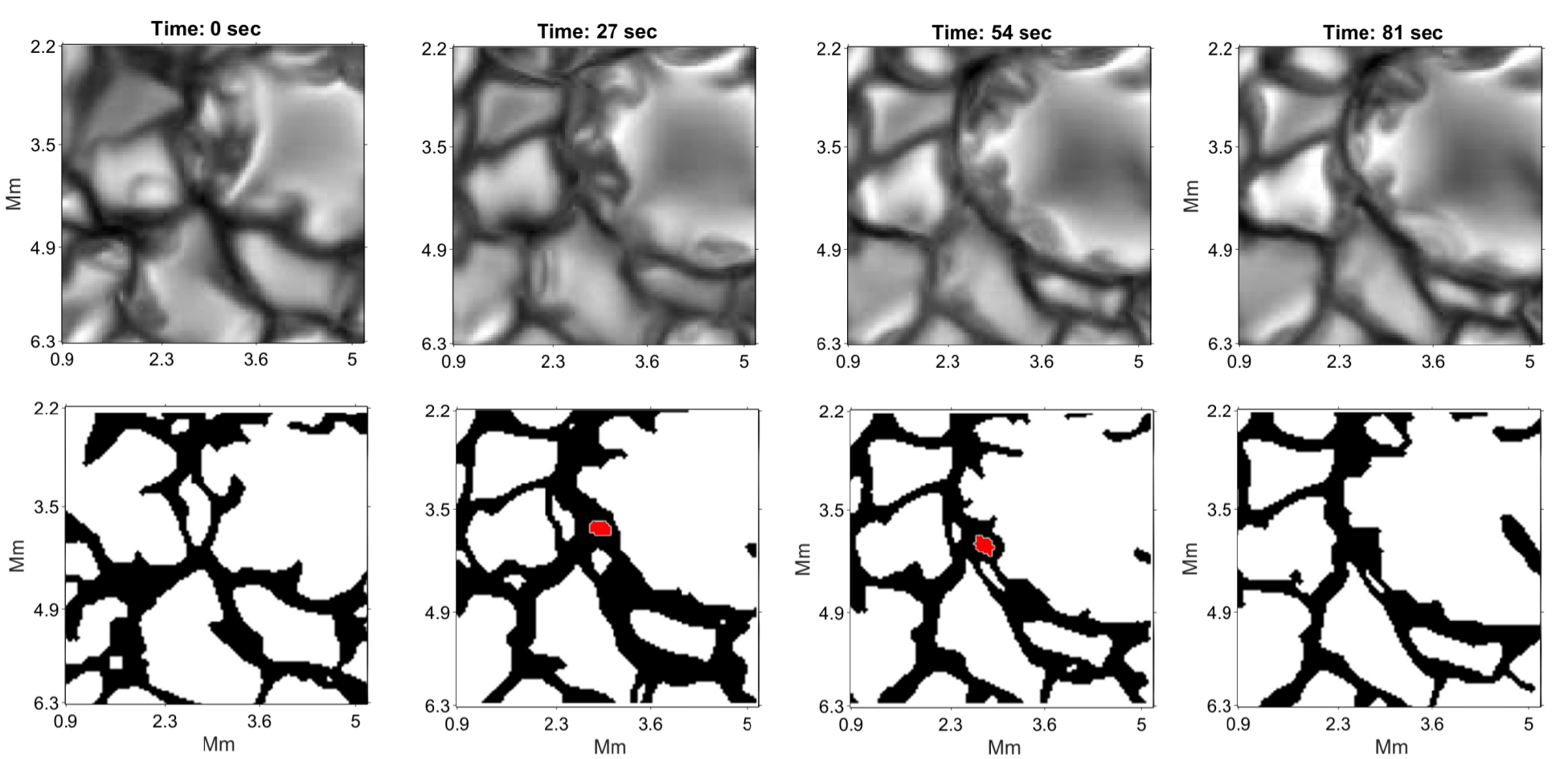}
                \caption{Application of the tracking algorithm applied to segmented profiles from the ANTARES simulation. The first row shows intensity maps of the evolution of a small granule, from its appearance to its dissolution. The second row illustrates the respective segmented intensity maps. The detected small granule is highlighted in red. This example shows that small and faint cells are detected, which is one of the advantages of the segmentation algorithm and the use of the vertical velocity in addition  to intensity maps.}
        \label{track}
\end{figure*}

Figure~\ref{track} illustrates the evolution of a small granule in the simulation. The images show the granules of interest in red. Each granule represents a tree which is parsed in order to carry out the lifetime and fragmentation analysis. 

\section{Results}
\label{sec:Results}

\subsection{Dynamics of granules} 
In this section we analyze the lifetime, size evolution, emergent intensity, vertical velocity and the variation of the position of granules in the photosphere / upper convection zone using ANTARES data and quiet Sun observations from the SUNRISE telescope. The simulation data consist of 900 time steps (2.25 h) and a total number of 10320 tracked granules. The SUNRISE data set consists of two time series with 100 images (50 minutes) and 9803 tracked granules. For the tracking we considered only those granules born and then dead within the time series. \\
 
Physical properties of the simulation, such as the vertical velocity, the emergent intensity and the geometrical depth of the segmented granules are evaluated on surfaces of the optical depth where $\tau$=1. This level of reverence is derived by using the mean Rosseland opacity. As a sort of an harmonic mean the Rosseland opacity gives more weight to the opacity contributions at wavelengths where the material is more transparent. The emerging intensity of the segmented granules is analyzed at the top of the computational box.  This can be understood as the intensity that reaches a virtual observer’s telescope. Because we used the gray approximation, the wavelength is irrelevant.

\subsubsection{Lifetime and fragmentation}
Our definition of the lifetime of granules is closely related to the definition given by \citet{1999ApJ...515..441H}: A granule starts its life either by fragmentation or appearance and dies by fragmentation or fading away. However, in our analysis we assume that a granule lives on if one fragment maintains a size of more than 80\% of the original granule. Granules evolve within the time series; hence, it is not possible to determine the age of granules detected in the first image of a time series. These granules are therefore excluded from the analysis.\\

\begin{figure*}
        \centering
                \includegraphics[width=1\textwidth]{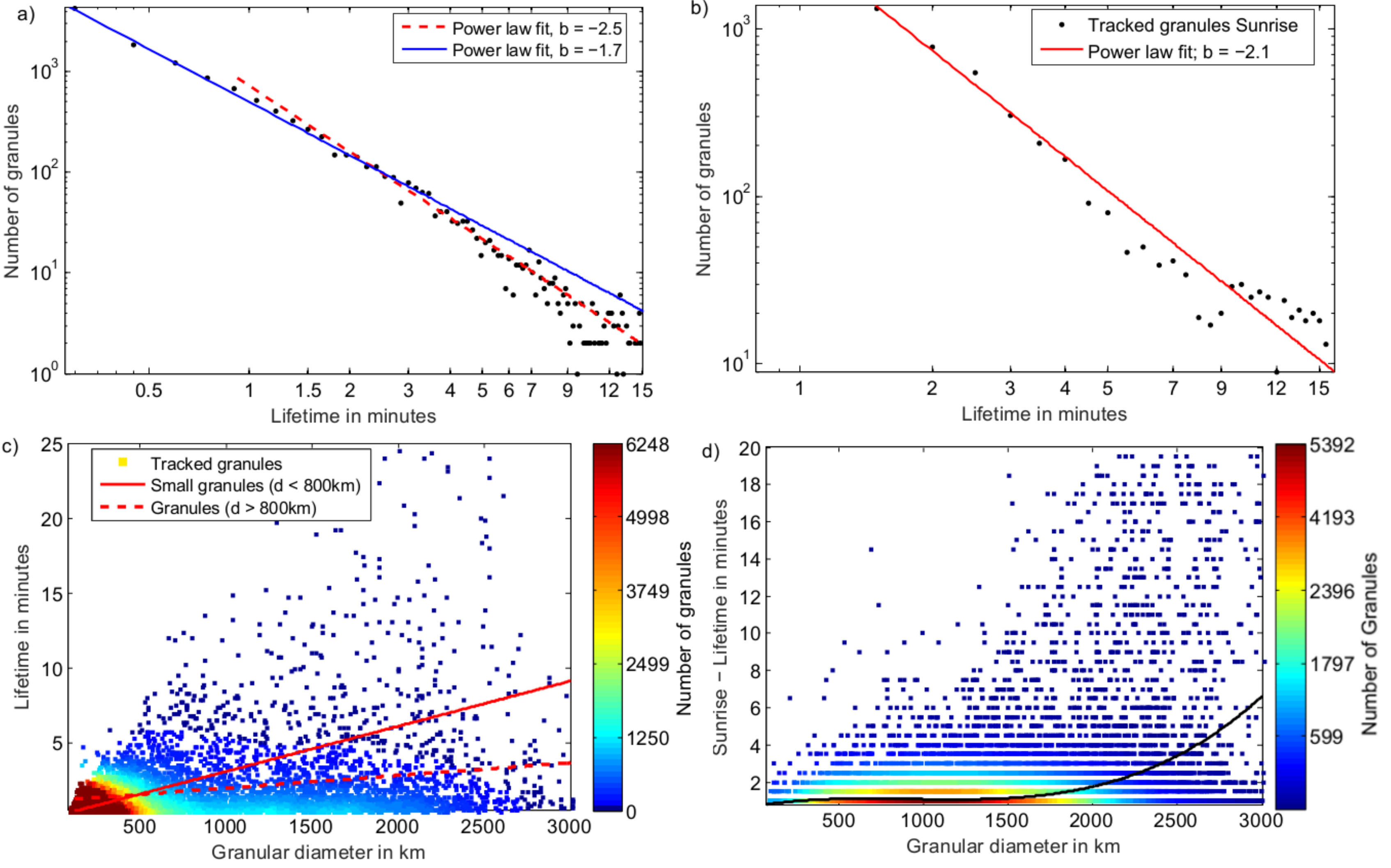}
                \caption{(a) Histogram of the lifetime of granules in the RHD simulation. (b) Histogram of the lifetime of granules in SUNRISE observations. (c) Density scatter plot of granular diameter vs. lifetime of tracked granules in simulations; (d) Density scatter plot of granular diameter vs. lifetime of tracked granules in SUNRISE observations; red indicates a high density in tracked granules.}
        \label{lifetimeRHD}
\end{figure*}

In Fig.~\ref{lifetimeRHD}a and Fig.~\ref{lifetimeRHD}b we show the lifetime histograms for the granules. The slopes of the histogram follow a power law: \\
\\ $f(t)=a \cdot t^{b}$\\ \\
where t is the lifetime and b the index. The lifetime of granules in the simulations that live for less than two minutes is distributed as a power law with an index b of -1.7. The distribution of granules that live longer than 2 minutes can be fitted by a power law with an index of -2.5. The lifetime of all tracked granules in the observations is presented by a power law with an index of -2.1.\\   
Figures~\ref{lifetimeRHD}c and~\ref{lifetimeRHD}d show the lifetime of granules depending on their maximum diameter. The density plot reveals that the majority of granules have diameters less than 500 km (illustrated in red). The linear fit (solid red line) shows an increase of granular size with time. Approximately 85\% of small granules have a lifetime of 2.5 minutes. Figure~\ref{lifetimeRHD}d shows that the majority of granules in observations have diameters between 500 and 1800 km. The black solid line indicates an increase of lifetime with size. \\ \\

\begin{figure*}
        \centering
                \includegraphics[width=1\textwidth]{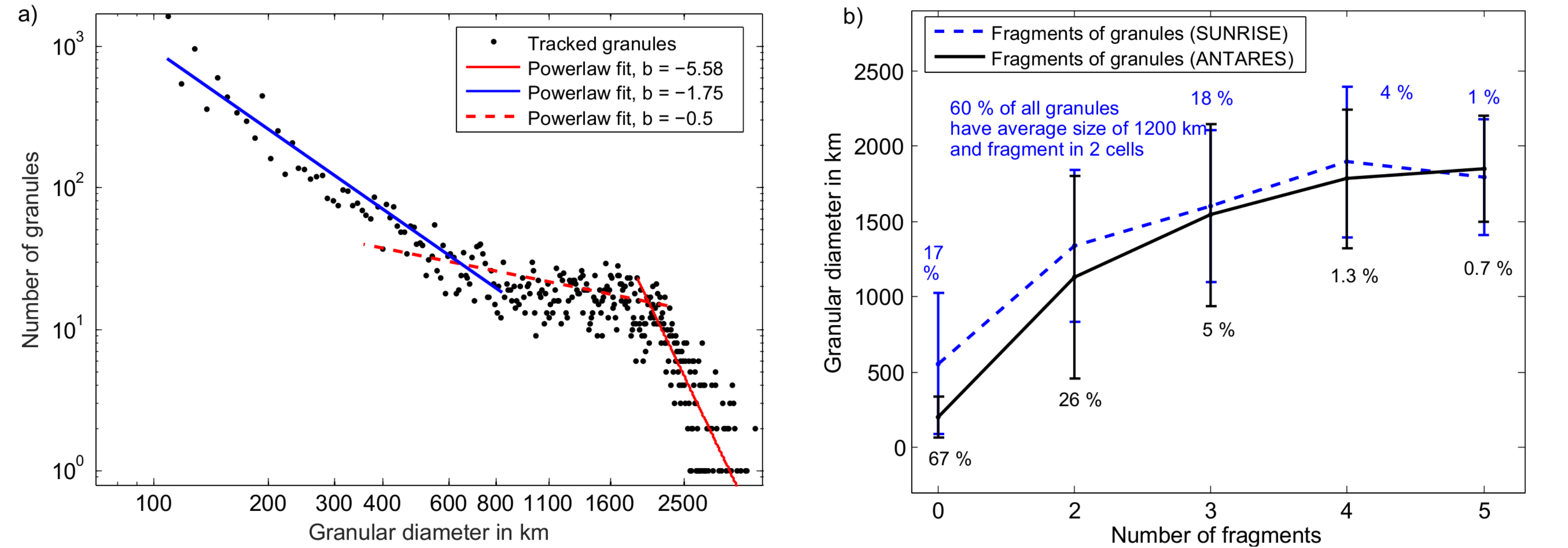}
                \caption{(a) Histogram of the diameter of granules in the simulation. The distribution is fitted by power law function with different indices. (b) Histogram of the the granular diameter vs. the number of fragments including the standard deviation displayed as error bars for granules in simulations (black solid line) and observations (dashed blue line). The percentage of granules that fragment into different cells is stated above and below the error bars.}
        \label{FragmentationRHD}
\end{figure*}

Figure~\ref{FragmentationRHD}a shows the histogram of the diameter of granules in simulations. The existence of two populations of granules is suggested by the changing slope of the histogram. A power law with an index of -1.75 can be fitted to granules with diameters less than 800~km, the slope changes for granules with diameters of between 800 and 1700 km (b = -0.5) and for granules with diameters larger than 1700~km (b = -5.58). \\

In our definition of the lifetime of granules we pointed out that granules die by fragmentation or by fading away. We analyze the fragmentation behavior of granules by determining the number of fragments for a given size of granule. Figure~\ref{FragmentationRHD}b shows the number of fragments depending on the diameter of the granule at the point of fragmentation. The solid black line shows that 67~\% of all granules in simulations have diameters smaller than 400 km and die without fragmentation. These granules have a mean diameter of 200 km and a lifetime of less than two minutes. 26~\% of all tracked granules fragment into two cells and 7~\% of all granules have a mean diameter of $\sim$1500~km and split into three or more fragments. The overall number of fragments rises with increasing granular diameter. The dashed blue line shows the fragmentation of granules in observations. In contrast to the non-fragmenting population in simulations, only approximately 17\% of granules in observations do not fragment. These granules have a mean diameter of 500 km and a larger size variation than in simulations. Approximately 60~\% of all tracked granules have an average diameter of 1200 km and fragment into two cells, 18\% fragment into three cells (d $\sim$1500~km), 5\% fragment into four and five cells (d $\sim$ 1700~km, 1800~km) and 17\% of granules (d $\sim$ 500 km) disappear without fragmentation.\\

\subsubsection{Evolution of structural properties of granules}
\begin{figure*}
        \centering
                \includegraphics[width=1\textwidth]{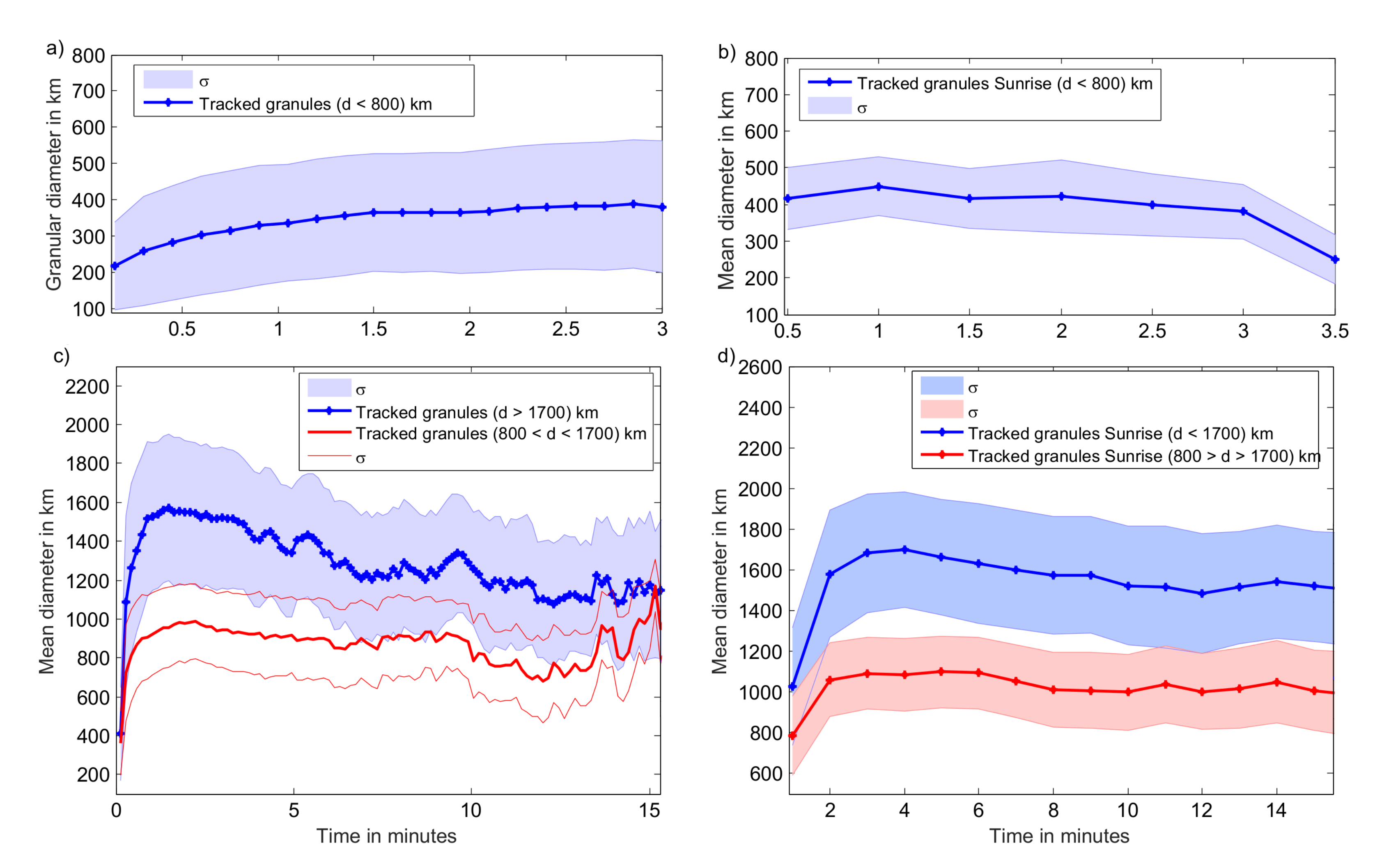}
                \caption{The curves represent the mean diameter of granules at a certain time instance: (a) Evolution of the diameter of small granules detected in the simulations with time. (b) Evolution of the diameter of small granules detected in SUNRISE IMaX observations. (c) The evolution of the diameter of large (> 1700 km; blue line) and medium-sized (800 km < d < 1700 km; red line) granules in simulations with time. (d) The evolution of the diameter of large (> 1700 km; blue line) and medium-sized (800 km < d < 1700 km; red line) granules in observations with time. The error bars of the curves represent the standard deviation with a confidence interval of 1-$\sigma$.}
        \label{diameterRHD}
\end{figure*}

Figure~\ref{diameterRHD}a shows the evolution of the diameter of detected small granules in the simulation. During the evolution, the small granules grow on average for 1.5 minutes and then remain constant in size. The majority of small granules (98\%) disappears after 3 minutes. The evolution of the diameter of small granules in observations (see Fig.~\ref{diameterRHD}b) indicates a slight decrease in size from $\sim${400} km to $\sim${250}~km after 2 minutes of lifetime. The standard deviation with a confidence interval of 1-$\sigma$ of the plots is given in shaded blue.\\
 
Figure~\ref{diameterRHD}c displays the evolution of medium-sized and large granules with time. Large granules (blue line with cross markers) reach their maximum size between one and two minutes and shrink slightly afterwards. The strong decrease in size in the first five minutes may be the result of a fragmentation process in at least two fragments. At a lifetime of 12 minutes the diameter decreases to a size of 1100 km and varies slightly until the cells disappear. Approximately 98\% of granules have a lifetime shorter than 15 minutes. The solid red line represents the lifetime of medium-sized granules. In the first ten minutes the group of medium-sized granules follows a similar trend compared to the large granules. Within one minute of their birth the granules reach an average size of 1000 km. The average diameter of the granules slightly decreases and increases again at a lifetime of 12 minutes. Medium-sized granules reach a mean size of $\sim${1000}~km at the end of their lifetime. \\

Medium-sized granules in observations (see Fig.~\ref{diameterRHD}d) reach an average maximum diameter of 1100 km after one minute, which decreases and reaches $\sim${1000} km at the end of their lifetime. In the first two minutes, large granules show a strong increase in size and reach a maximum diameter of 1500~km until the end of their lifetime.\\

\begin{figure*}
        \centering
                \includegraphics[width=1\textwidth]{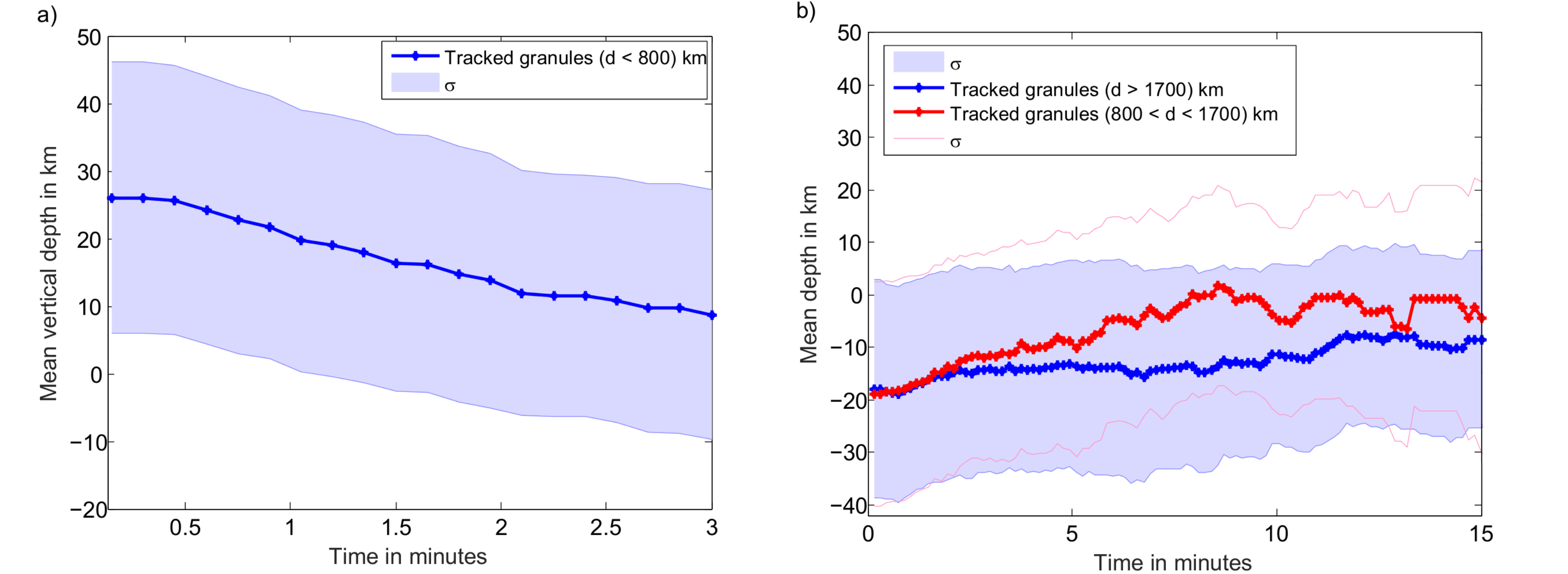}
                \caption{(a) Evolution of the location of small granules in the simulation box with time. The mean geometrical depth is used for determining the location. (b) The evolution of the depth of large (> 1700 km) and medium-sized (800 km < d < 1700 km) granules with time.}
        \label{depthRHD}
\end{figure*}

In paper I we showed that small granules are located deeper into the convection zone based on the mean geometrical depth at the $\tau_1$-iso-surface, which is the calculated surface of the optical depth where $\tau$=1. The evolution of the vertical location of granules in the simulation box during their lifetime is illustrated in Figure~\ref{depthRHD}. At the point of their detection they are located 26~km below the $\tau_1$-iso-surface. The positive numbers indicate the locations below the $\tau_1$-iso-surface and the negative numbers, above the surface. At a lifetime of 30 seconds small granules start to rise and reach their minimum depth at 11.5~km below the surface at the end of their lifetime. During their lifetime, the tracked small granules are found to be located below the iso-surface.\\
 
Figure~\ref{depthRHD}b illustrates the evolution of the position in the simulation box for large and medium-sized granules. Both groups of granules are detected above the calculated surface. They start to sink shortly after their detection and reach the height of the average optical surface at the end of their lifetime. In contrast to large granules, which are located above the calculated surface throughout their entire lifetime, medium-sized granules show up and down movements near the level of the calculated surface at a lifetime of eight minutes. Large granules show a less pronounced up and down movement during their evolution.

\subsubsection{Evolution of physical properties of granules}

\begin{figure*}
        \centering
                \includegraphics[width=1\textwidth]{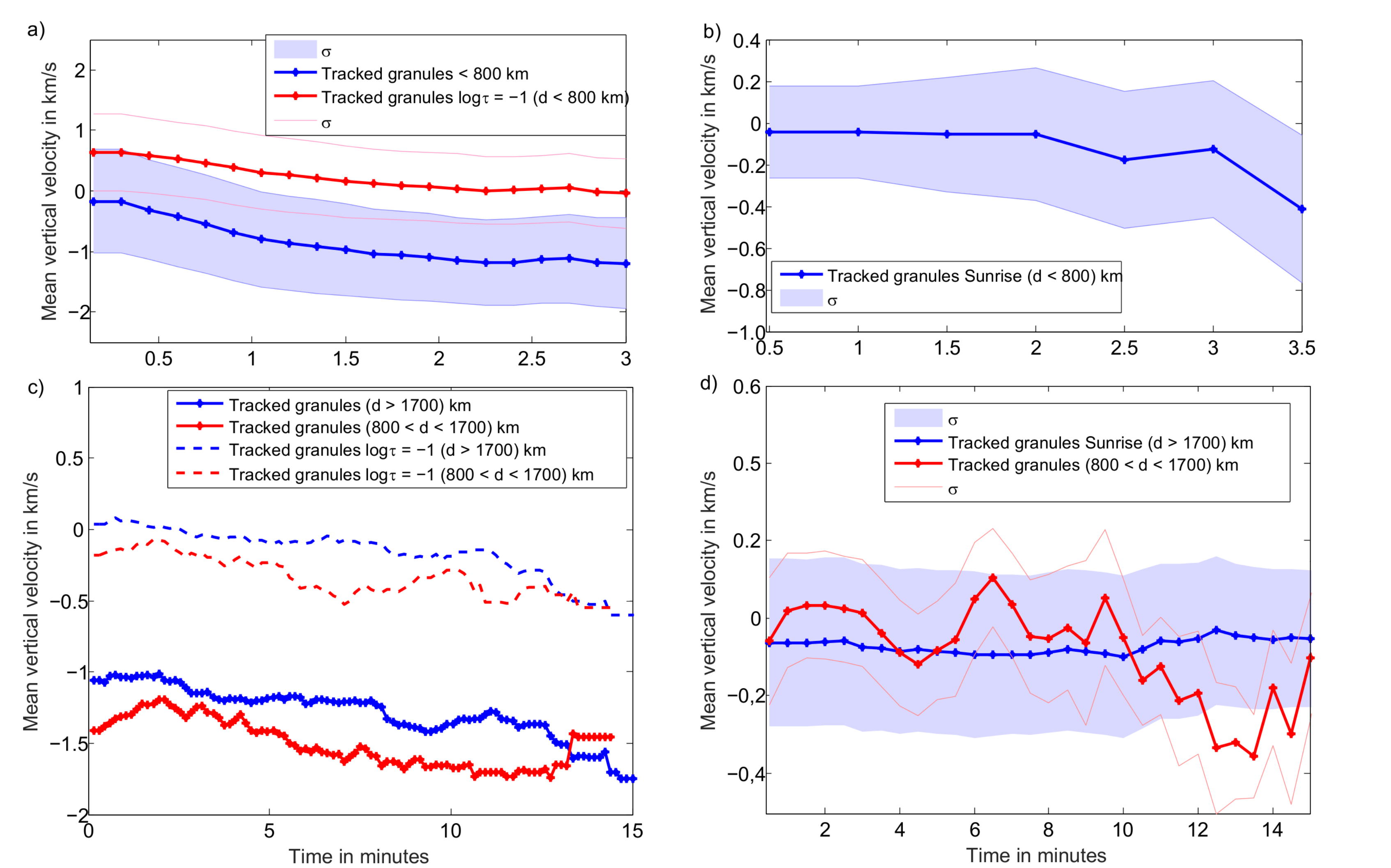}
                \caption{(a) Evolution of the velocity of small granules in the simulation at the $\tau_1$-iso-surface (blue line) and the log~$\tau_{-1}$-iso-surface (red line). (b) Small granules in observations. (c) The evolution of the velocity of large (> 1700 km; blue) and medium-sized (800 km < d < 1700 km; red) granules in the simulation at the $\tau_1$-iso-surface (solid line) and the log~$\tau_{-1}$-iso-surface (dashed line) with time. (d) The evolution of the vertical velocity of large granules (> 1700 km; blue line) and medium-sized (800 km < d < 1700 km; red) granules with time in observations. The error bars of the curves represent the standard deviation with a confidence interval of 1-$\sigma$.}
        \label{velocityRHD}
\end{figure*}

The evolution of the vertical velocity of detected granules is shown in Figure~\ref{velocityRHD}. The velocity derived from the inversions is formed at an optical depth where the line is most sensitive to the velocity \citep[see e.g.][]{2005A&A...439..687C}. The Fe I 525.02 nm line observed by IMAX is a photospheric line, formed above the $\tau_1$-iso-surface. Hence, we calculated the average velocity of granules at the log $\tau_{-1}$-iso-surface in the simulations, which is situated approximately 80 km above the $\tau_1$-iso-surface, to compare the results of the velocity evolution from simulations to observations.\\  

The small granules (see Fig.~\ref{velocityRHD}a) in simulations at the log $\tau_{-1}$-iso-surface show a constant downflow of 0.5 km/s with a slight increase during their evolution. Small granules at the $\tau_1$-iso-surface exhibit a slight upflowing motion ($\sim${0.06} km/s) shortly after their detection, which increases during their lifetime to an upflow velocity of 1.1 km/s until they fade away. This evolution is also visible for small granules in observations (Fig.~\ref{velocityRHD}b) that reach an upflow velocity of 0.4 km/s at the end of their lifetime. Their evolution is closer to the vertical velocity evolution of small granules in simulations at the $\tau_1$-iso-surface. Figure~\ref{velocityRHD}c shows the evolution of the vertical velocity for medium-sized and large granules at both iso-surfaces. Both groups of granules exhibit upflows during their evolution, which is less pronounced in the higher layers. Large granules show an increase in upflow velocity starting from -1 km/s at the beginning of their lifetime to -1.7 km/s at the end of their lifetime. After the detection of medium-sized granules (red line) the vertical velocity decreases and starts to increase again at a lifetime of 2 minutes to reach a local maximum of $\sim${-1.75} km/s at 13 minutes. At a lifetime of 14 minutes the average vertical velocity of medium-sized granules decreases to a fade-out velocity of -1.2 km/s and of -0.5 km/s in the higher atmosphere. The values at the log~$\tau_{-1}$-iso-surface of these groups of granules is in good agreement with the observations (see Fig.~\ref{velocityRHD}d). The decrease of the vertical velocity is also visible at a lifetime of 14 minutes. Large granules, on the other hand, show a stable upflow velocity of $\sim$0.05 km/s.\\

\begin{figure*}
        \centering
                \includegraphics[width=1\textwidth]{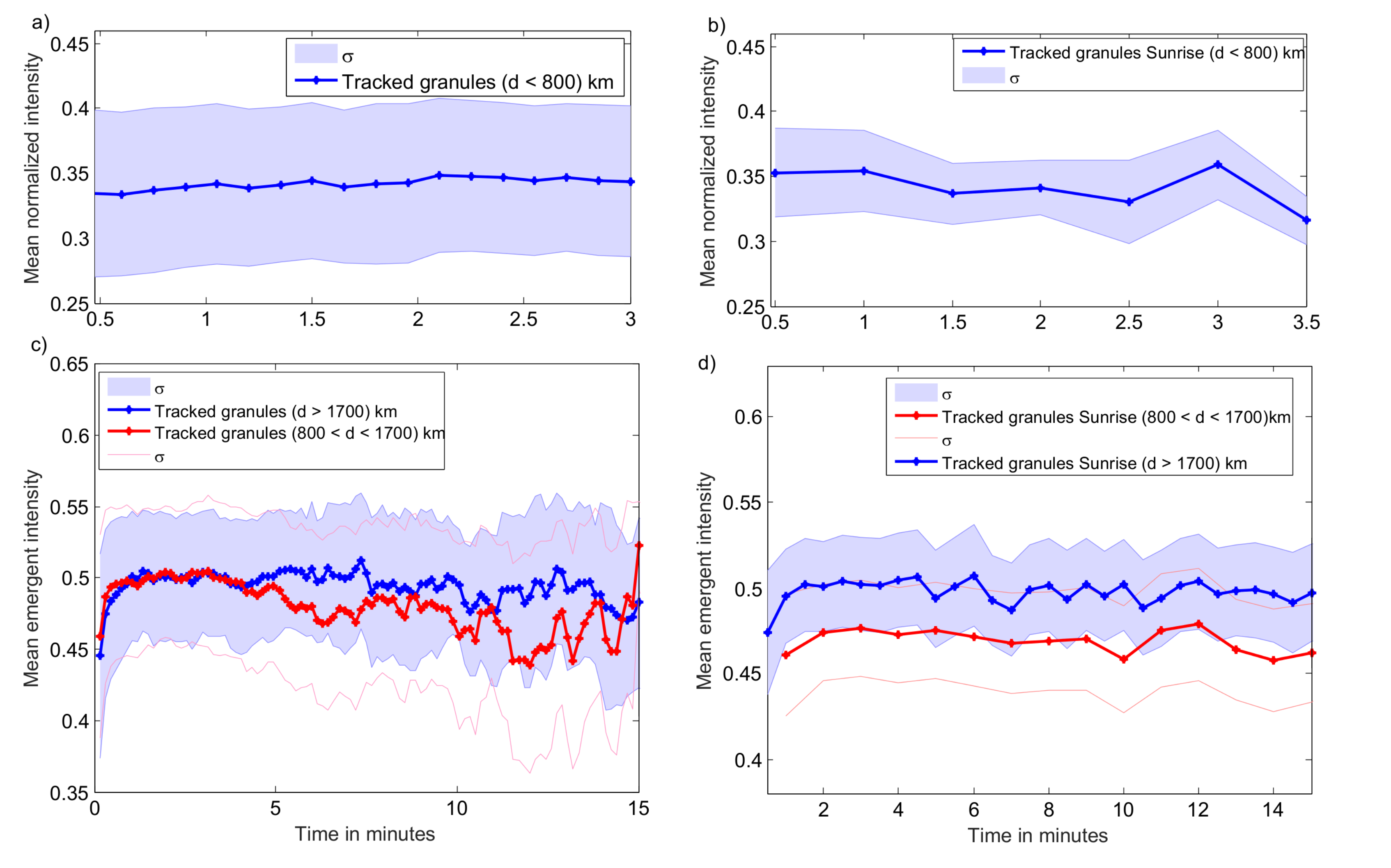}
                \caption{(a) Evolution of the emergent intensity of small granules in simulations. (b) Evolution of the intensity of small granules tracked in SUNRISE IMaX observations. (c) The evolution of the emergent intensity of large (> 1700 km; blue line) and medium-sized (800 km < d < 1700 km; red) granules in simulations with time. (d) The evolution of the emergent intensity of large (> 1700 km; blue line) and medium-sized (800 km < d < 1700 km; red) granules with time in observations. The error bars of the curves represent the standard deviation with a confidence interval of 1-$\sigma$.}
        \label{intensityRHD}
\end{figure*}

The evolution of the emergent intensity of small granules shows a constant intensity of 0.34 (see Fig~\ref{intensityRHD}a) and, for small granules, a slight decrease in brightness  in observations at a lifetime of approximately three minutes to an average intensity value of 0.33 (see Fig~\ref{intensityRHD}b). The mean normalized emergent intensity of medium-sized and large granules in simulations is shown in Figure~\ref{intensityRHD}c. The intensity of medium-sized and large granules in simulations strongly increases in the first minute to an average intensity of 0.5. At a lifetime of 12 minutes medium-sized granules reach a global minimum of 0.45 followed by an increase to a mean intensity of 0.5. The variations in the mean emergent intensity in medium-sized granules starting from minute 12 until the end of their lifetime is due to the reduced sample statistics and can be neglected. Large granules show a slight decrease in intensity during their lifetime. In observations, medium-sized granules  (see Fig~\ref{intensityRHD}d) show only minor variations during their evolution towards the end of their lifetime. Large granules (blue line) show an increase in brightness in the first minute of lifetime to an intensity value of 0.5, which remains almost constant throughout their entire lifetime.

\subsection{Dynamics of Vortices in simulations}
The dynamics of granules are influenced by surrounding magnetic flux concentrations and convective motions such as strong downflows or vortex flows. Of particular interest are the factors, which influence the dynamics of small granular cells, as well as the mechanisms, which contribute to their generation and influence their lifetime. The study of the convective motion that surrounds the small cells provides valuable information about their evolution. We use the numeric simulations to analyze the vortex motions in the vicinity of granules. On the one hand, we computed the horizontal and vertical components of the vorticity in the RHD data at the $\tau_1$-iso-surface and segmented the vortices based on the strength of the swirls. On the other hand we detected and tracked strong horizontal vortices in the 2.5 h simulation run, that are visible as small-scale bright features (referred to as convective bright points (CBPs)). We applied the tracking algorithm to the 1400 detected features and analyzed their dynamics. \\

\begin{figure*}
        \centering
                \includegraphics[width=0.8\textwidth]{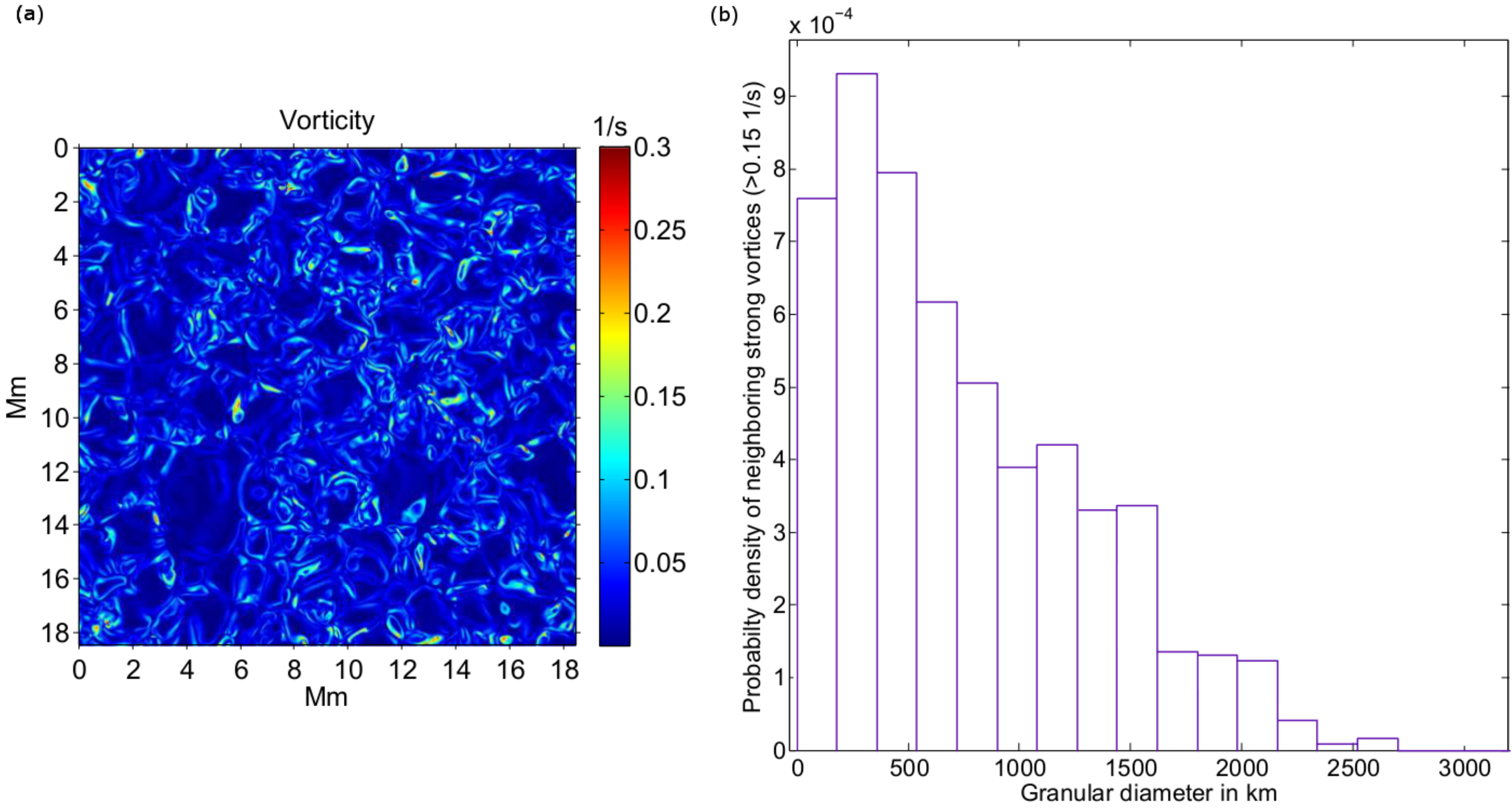}
                \caption{(a) $\tau_1$-iso-surface of the horizontal component of the vorticity. Strong vortex motions are colored red. (b) Probability density functions of the neighboring strong vortices.}
                        \label{HorizontalVortex}
\end{figure*}

\begin{figure*}
        \centering
                \includegraphics[width=1\textwidth]{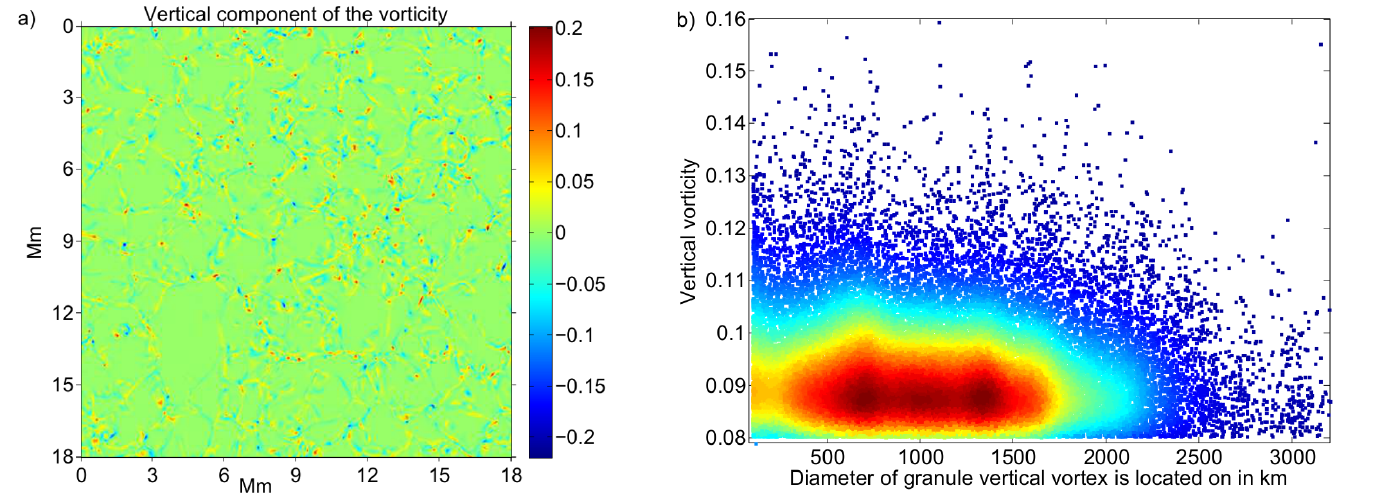}
                \caption{(a) Snapshot of the vertical component of the vorticity; strong vortices are colored in red and blue. (b) Density scatter plot of the strength of the vertical component of the vorticity vs. the diameter of a granule, which is located at the position of a detected vertical vortex; red indicates a high density.}
        \label{VerticalVortex}
\end{figure*}

Figure~\ref{HorizontalVortex}a shows strong horizontal components of the vorticity in the simulations (displayed in red/yellow), which are found to be located at the vertices of granules as well as at the location of small granules. In Figure~\ref{VerticalVortex} a strong vertical vorticity is located in the intergranular lanes and marked in red (upwards directed) and blue (downwards directed).\\

We analyzed the RHD data by applying the segmentation masks, which were computed using the segmentation algorithm, to study the neighborhood of each detected granule in each time step. Strong vortices ( > 0.08 1/s) in the vicinity of granules (100 km border to center distance) are taken into account.\\
Figure~\ref{HorizontalVortex}b shows the probability of a distinct size of granule to be surrounded by strong horizontal vortices. The probability density function shows that strong vortices are detected in the neighborhood of granules with a diameter of $\sim${500}~km. The shape of the histogram depicts the turbulent character of small granules and shows that the dynamics of small granules at the surface are influenced by the horizontal vortex tubes.\\

Figure~\ref{VerticalVortex}b is a density plot that shows the strength of the vertical component of the vorticity as a function of the size of a granule, at the location of a vertical vortex. The shape of the plot shows that the strongest vertical vortices are located at the location of small granules ( d < 800 km). Our result shows that strong vertical vortices are located at the position of granules in a diameter range of 500~km to 1600~km. Two distinct concentrations are visible at 700 km and 1400 km.\\

\begin{figure*}
        \centering
                \includegraphics[width=0.8\textwidth]{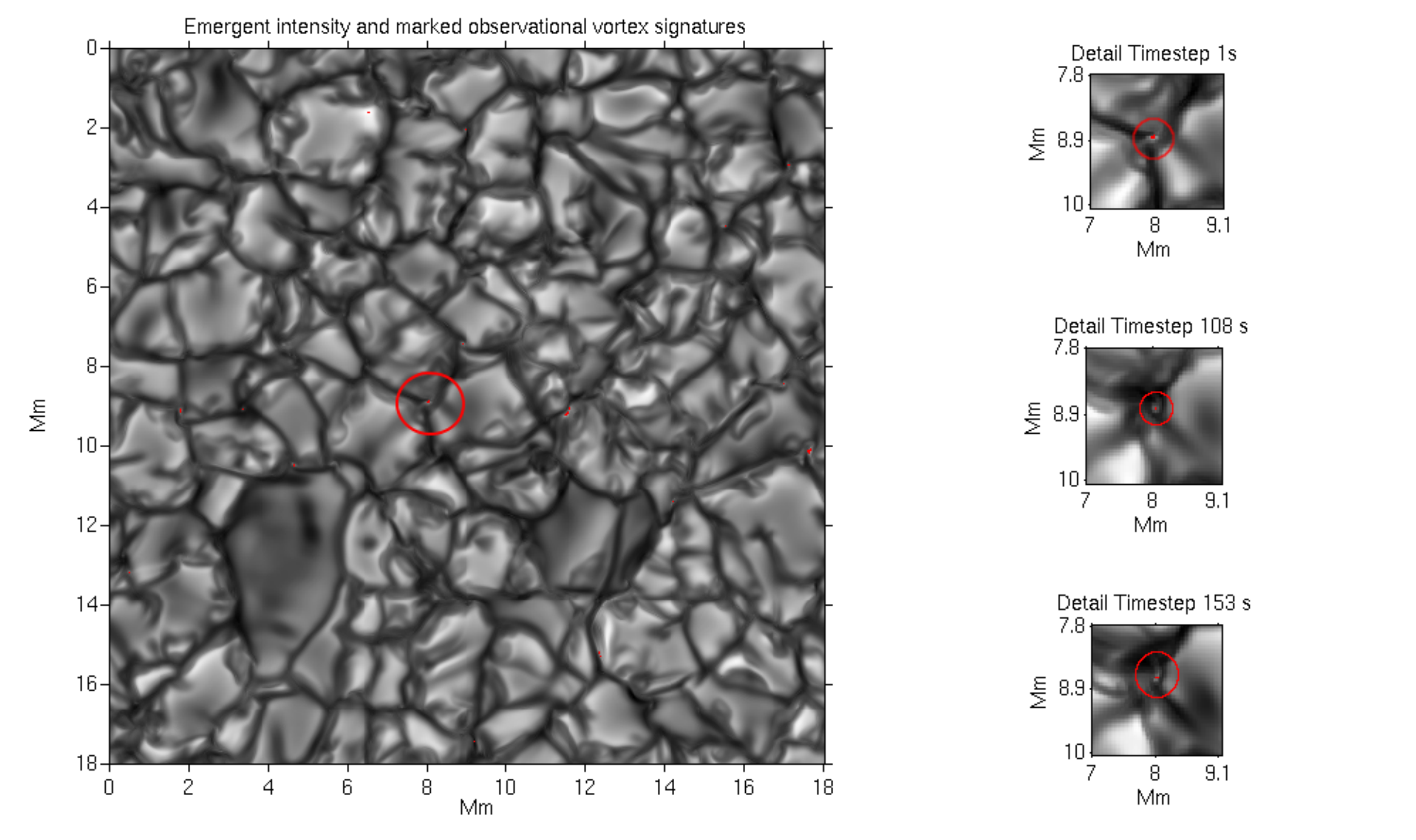}
                \caption{The left panel shows an emergent intensity snapshot of the simulation run. Detected CBPs are marked in red. The panels at the right side illustrate the evolution of the detected CBPs, which is encircled in red in the intensity snapshot (left panel).}
                        \label{CBPTimeSeries}
\end{figure*}

Figure~\ref{CBPTimeSeries} shows the detected CBPs in an emergent intensity snapshot of the simulation. The features detected are marked in red. These CBPs are characterized by their high intensities and most of all by their high contrast in comparison with the surrounding granules. We identified these structures by applying a global intensity threshold and a highpass filter in the frequency domain of the intensity maps. In the right panels of this figure we demonstrate the evolution of the CBP marked with a circle in the intensity profile (lifetime: 170 s). The convective features apparently decrease in brightness during their evolution.\\

\begin{figure*}
        \centering
                \includegraphics[width=1\textwidth]{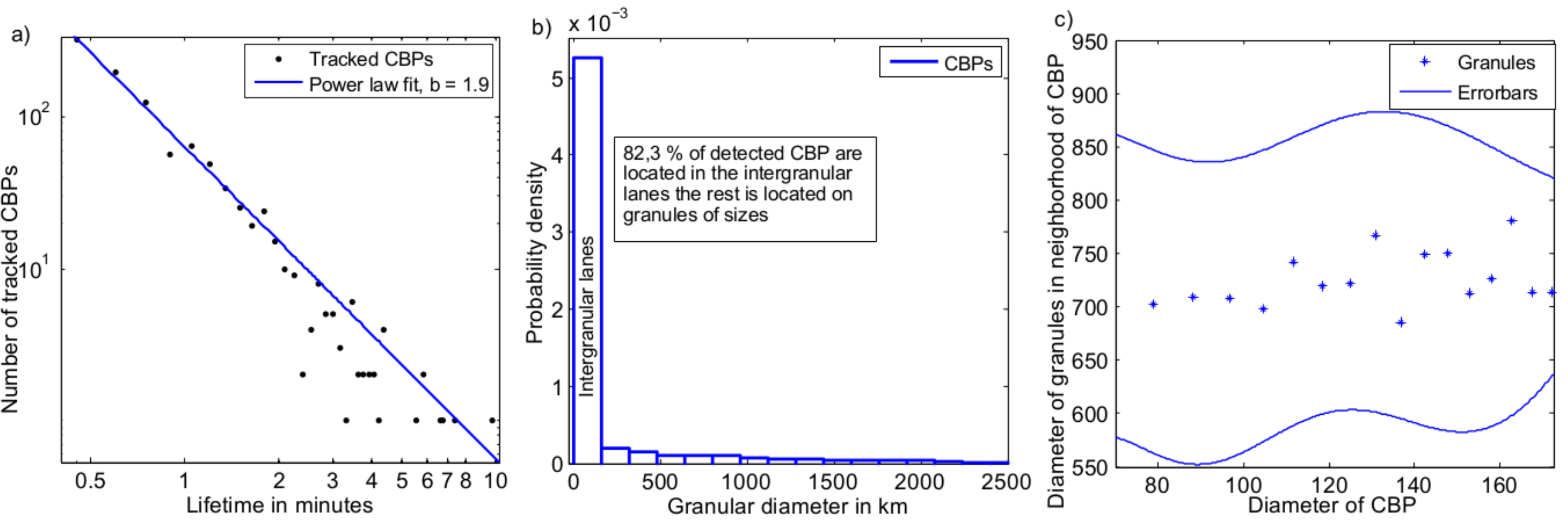}
                \caption{(a) The histrogram of lifetimes of CBPs. (b) Probability density of detected CBPs located at the position of detected granules of differing size. (c) Diameter of CBPs vs. the diameter of neighboring granules including the standard deviation (1-$\sigma$).}
        \label{CBP1}
\end{figure*}

The histogram of the lifetime of granules (see Fig.~\ref{CBP1}a) can be fitted by a power law with an index of b = -1.9. We found that the majority of features have lifetimes up to seven minutes. The location of the segmented convective bright points is analyzed in Figure~\ref{CBP1}b. The histogram shows a 82.3~\% probability that these vortex motions are found in the intergranular lanes. Only a fraction of 17.7~\% of CBPs is located at the position of a detected granule. These findings lead to the study of the neighborhood of the segmented bright structures. From the analysis in Figure~\ref{CBP1}c it can be concluded that bright features are predominantly surrounded by small granules. The average diameter of the CBPs is $\sim$ 80 km.\\

\begin{figure*}
        \centering
                \includegraphics[width=1\textwidth]{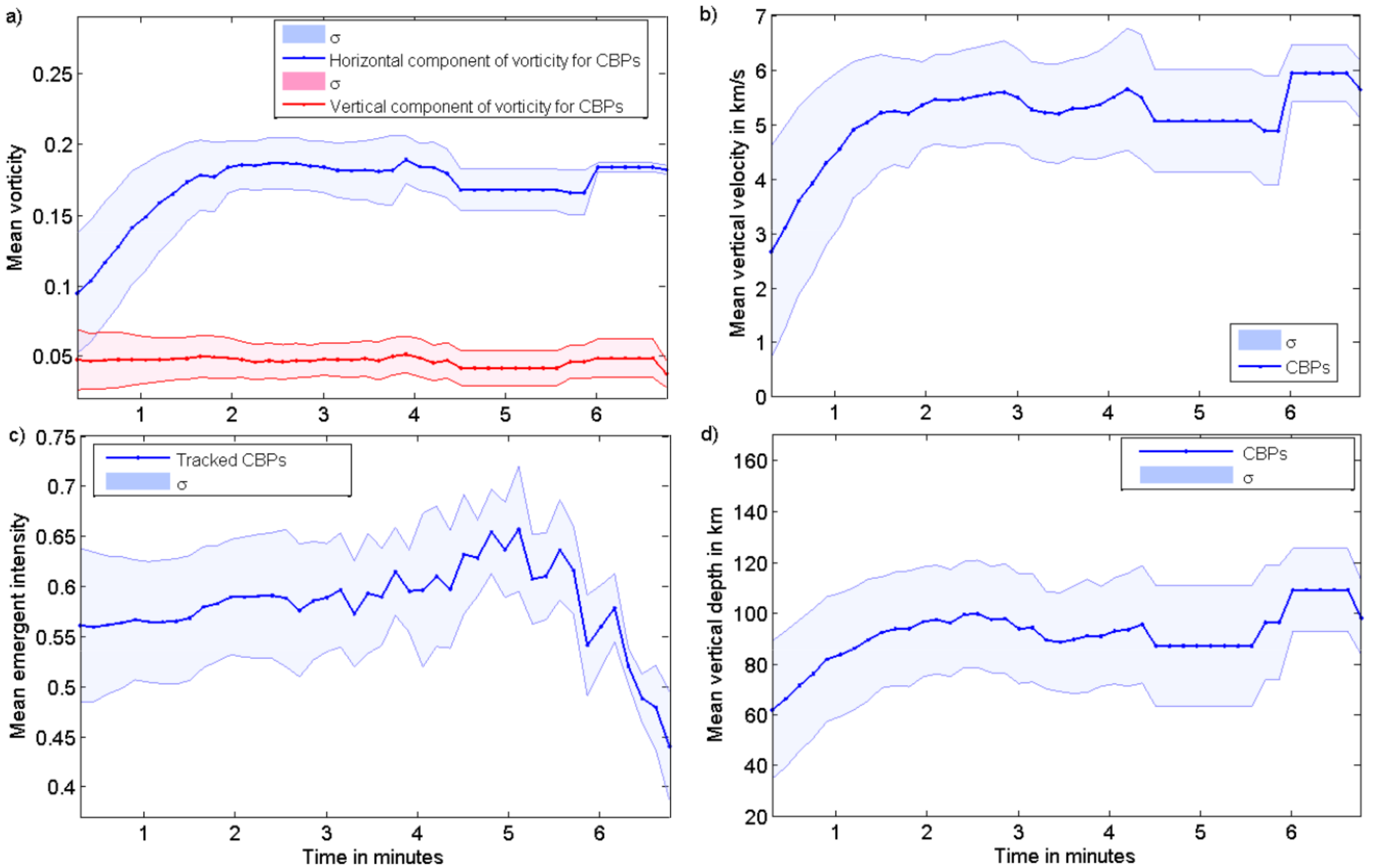}
                \caption{(a) The evolution of the vertical (red line) and horizontal (blue line) components of the vorticity of CBPs. (b) The mean vertical velocity evolution during the lifetime of CBPs. (c) The emergent intensity evolution of CBPs. (d) Evolution of the location of CBPs in the simulation box with time. The error bars of the curves represent the standard deviation with a confidence interval of 1-$\sigma$.}
        \label{CBP2}
\end{figure*}
 
Figure~\ref{CBP2} illustrates the evolution of physical properties of tracked CBPs with time. The evolution of the horizontal and vertical component of the vorticity of CBS (see Figure~\ref{CBP2}a) shows an enhancement of the horizontal component of the vorticity of CBPs with lifetime. The averaged vertical velocities show strong downflows reaching values up to 6 km/s at a lifetime of six minutes. The mean emergent intensity of the tracked CBPs (see Fig~\ref{CBP2}c) increases during their evolution and strongly decreases at a lifetime of five minutes. CBPs are predominately found in the intergranular lanes located deeper in the convection zone. The evolution of the position of the CBPs in the simulation box shows that after their detection they start to sink $\sim$ 100 km below the averaged level of the surface.

\section{Discussion and conclusions}
\label{sec:discussion}
The high-resolution RHD simulations carried out with the ANTARES code allow us to infer characteristic physical quantities for convective phenomena, on any layer in the box. In this study we present a newly developed temporal tracking routine for the solar granulation based on the results obtained from the application of the segmentation algorithm to simulation profiles. Tracking of the granules results in a tree-like structure for each granule. This allows a determination of the lifetime of granules detected in ANTARES simulations and SUNRISE IMaX observations and a tracking of the evolution of physical quantities. To study their influence on the dynamics of granules, we investigated the vertical and horizontal vortex motions in the vicinity of granules. We analyzed vortex motions based on the strength of the swirls and on their characteristic appearance in emergent intensity profiles.\\ 

In our study of the lifetime of granules in simulations and observations we detected a high number of tracked granules with a lifetime of less than two minutes. The lifetime distribution of granules in simulations can be fitted by a power law with different indices. Granules with lifetimes shorter than two minutes (small granules) are distributed as a power law with an index of -1.7. The existence of two populations of granules is also suggested by the changing slope of the histogram of diameters of granules in simulations. These results are in accordance with observations in \citet{2012ApJ...756L..27A}.\\

The analysis of the fragmentation of granules in the simulations showed that 67~\% of detected granules have a mean diameter of 200 km and die without fragmentation. The number of fragments rises with increasing granular diameter.  In their MHD simulations, \citet{2013A&A...558A..49B} label the small cells as erroneous detections. In paper I we showed that small granules strongly contribute to the total number of detected granules, and hence, have a major influence on the dynamics of the population of all granules.  Approximately 83 \% of granules tracked in observations fragment into two or more cells, which is in contrast to the fragmentation behavior of granules in the simulation. This divergence in the fragmentation process may be a direct consequence of the high number of detected small granules in the simulations. In contrast, the majority of granules detected in observations have diameters between 500 km and 1500 km. The difference in number of detected small granules may be due to the presence of the magnetic field, which fills the intergranular lanes and suppresses the upflow of the evolving small granules \citep[see e.g.][]{2009LRSP....6....2N}. An additional explanation for these diverging results may be the temporal resolution of 32 seconds of the observations, which is 3.5 times lower compared to the simulations. A granule that fragments  into more than two cells in the successive time step could fade away within the 30 seconds and a new granule could have appeared at the same location.\\

The evolution of the diameter of granules tracked in simulations and observations shows that small granules grow in size until they reach a lifetime of 1.5 minutes, and then slightly decrease. The diameter of small granules varies in a small range; hence, small granules do not develop into medium-sized granules. This may be due to the smaller pressure excess of the small granules that limits their growth and shrinks their size \citep[see e.g.][]{1989ApJ...342L..95S}. Small granules seem to appear predominantly in the intergranular lanes and disappear again after three minutes. The results of the diameter evolution in \citet{1989ApJ...336..475T}, \citet{1999ApJ...515..441H} and \citet{2004A&A...428.1007D} are comparable to the diameter evolution of small granules, which show an increase in size with time. We have shown that medium-sized and large granules behave differently. In the simulation and in observations, large granules  reach their maximum size within the first minutes of lifetime and then decrease in size. However, medium-sized granules tracked in simulations reach a local size minimum at a lifetime of 12 minutes and reach their maximum size at the end of their lifetime. This drop in diameter at a lifetime of ten minutes may be due to the fragmentation of the granule in a small and a large cell. The large cell, which must remain larger than 80\% of the original granule in order to still be considered as the same cell, slightly increases after 12 minutes of lifetime until it reaches a diameter of 1100 km. This reversal of size for medium-sized granules in simulations at a lifetime of 12 minutes is also visible in a reversal in the evolution of the velocity at 12 minutes. The analysis of the diameter evolution of granules does not show a clear difference in the size of granules between simulations and observations, which may be expected as a result of the influence of the magnetic field \citep[e.g.][]{1989ApJ...336..475T}. The quiet Sun observations are therefore comparable with RHD simulations. \\ 

The analysis of the variation of the vertical position of granules in the simulation reveals that small granules rise during their evolution and hold their position below the calculated averaged level of $\tau$=1. This is due to their location in the optically deeper intergranular lanes. Medium-sized granules sink until they reach the level of the surface and then rise slightly.\\

In simulations at the log $\tau_{-1}$-iso-surface and in observations, medium-sized granules show a decrease in vertical velocity which initiates the end of their lifetime. In simulations, these granules show an intensity decrease ten minutes after their birth. In contrast to medium-sized and large granules, the variation in intensity of small granules is an order of magnitude lower. This can be explained by their shorter lifetimes and their location below the calculated surface. The separation in small, medium and large granules reveals different intensity evolutions for the different groups of granules, which provides a more detailed picture in comparison to previous studies that use averaged values of granules \citep[see e.g.][]{1999ApJ...515..441H,2004A&A...428.1007D}  \\

We found that strong horizontal and vertical vortex motions have a high tendency to be located in the vicinity of small granules. Strong horizontal vortices are predominantly located in the intergranular lanes and may influence the lifetime of small granules. Horizontal vortex tubes were also detected in SUNRISE observations, located along the edges of granules \citep{2010ApJ...723L.180S}. The strongest vertical vortices are detected on small granules. This is in accordance with findings in \citet{2012PhyS...86a8403K} where vertical vortex tubes were found to be formed inside granules as a result of the local instability caused by small upflowing plumes. In the scatter plot, two concentrations of vertical vortices are visible at the location of granules with 700 km and 1400 km in diameter, further supporting  the idea of the existence of two populations of granules. Strong vortices are more likely to be located at the center of small granules than of medium-sized granules. Vortices on medium-sized granules are often the result of horizontal vortex tubes near the vertices of the granules that become vertical vortex tubes. This is due to the strong downflows resulting from angular momentum conservation \citep[see e.g.][]{1989ApJ...342L..95S}. \\
 
In movies of the emergent intensity profiles we detected vortex-like features, which appear bright and high in contrast compared to their surroundings. These convective bright points appear at the vertices of medium-sized granules and live for up to seven minutes until they disappear in the intergranular lanes. These results are in accordance with observations (average lifetime of five minutes in \citet{2008ApJ...687L.131B}). New studies into the lifetimes of vortices in observations carried out by \citet{2015SoPh..290..301V} indicate a lifetime of 15 to 19 minutes. The authors claim that the stability and lifetime of the vertical motions is strongly linked to the evolution of the granular pattern. The detected CBPs are generated at the vertices of the granules as a result of the conservation of angular momentum of the plasma that sinks down and then drifts into the intergranular lanes (bathtub effect; see e.g. \citet{1985SoPh..100..209N}), where they disappear as a result of the turbulent character of the surrounding downflows.\\
The histogram of the lifetime distribution of CBPs in simulations shows that the probability that these vortex motions are found in the intergranular lanes is 82.3 \%. Only 17.7 \% are located at the position of detected granules. The bright features located at the position of detected granules are of an elongated shape. These findings are possibly related to the horizontal vortex tubes observed as bright rims in Steiner et al. (2010).\\

The intensity of CBPs is 10 \% higher compared to the average brightness of granules. The brightness enhancement of CBPs is the result of a side-wall heating in the interior of the vortices \citep[see e.g.][]{2013A&A...558A..49B}. CBPs have a horizontal diameter of $\sim$ 80 km which is consistent with results in \citet{2011A&A...533A.126M}. These vortex motions are difficult to observe with current telescopes, therefore higher resolution observations and missions are needed. The analysis of the evolution of physical properties of CBPs shows strong vertical vortex motions. The investigation of movies shows that these features are generated at the vertices of the granules and move into the intergranular lanes. These small fading structures show high downflows ($\sim$ 6 km/s) and a high vertical component of the vorticity. 

\paragraph{Outlook: }
In a subsequent study, we will focus on the dynamics of small granules in MHD simulations and compare their statistical and physical properties to observations of the quiet Sun. Furthermore, we will analyze how the magnetic field influences the fragmentation process of granules and the evolution of the intensity.\\ 
The dynamics of vertical vortex tubes in 3D will be conducted in a future study on MHD simulations. Vertical vortex tubes are an important field of study as they may generate MHD waves or chromospheric tornadoes which may heat the Chromosphere after dissipation.

\begin{acknowledgements}
The authors thank the anonymous referee for constructive comments that helped to improve the manuscript and presentation of the results. The research work was funded by the Austrian Science Fund (FWF): P27765 and P23618. The model calculations have been carried out at VSC (project P70068, H. Muthsam). We thank Luis Bellot Rubio from the Instituto de Astrofisica de Andalucia (CSIC) supplying the inverted SUNRISE IMaX data used in this paper.  
\end{acknowledgements}
\bibliographystyle{aa}
\bibliography{literature}
\end{document}